# Observation of Joule-Thomson photon-gas expansion

Marco S. Kirsch[1,*], Georgios G. Pyrialakos[2,*,‡], Richard Altenkirch[1], Mahmoud A. Selim[2], Julius Beck[1], Tom A. W. Wolterink[1], Huizhong Ren[2], Pawel S. Jung[3], Mercedeh Khajavikhan[2], Alexander Szameit[1], Matthias Heinrich[1] and Demetrios N. Christodoulides[2,⁑]

[1] Institute of Physics, University of Rostock, Albert-Einstein-Str. 23, 18059 Rostock, Germany.

[2] Ming Hsieh Department of Electrical and Computer Engineering, University of Southern California, Los Angeles, CA 90089, USA.

[3] College of Optics & Photonics-CREOL, University of Central Florida, Orlando, FL 32816, USA.

[*] These authors contributed equally to this work.

[‡] email: gp_019@usc.edu [⁑] email: demetri@usc.edu

**Abstract:**

In recent years, a self-consistent optical thermodynamic framework has emerged that offers a systematic methodology to understand, harness and exploit the complex collective dynamics of multimode nonlinear systems. These developments now allow consideration of a series of longstanding problems in optics, including the prospect of funnelling the entire power flowing in a multimode system into its ground state, for which no methodology currently exists. Here, we demonstrate an all-optical Joule-Thomson expansion process mediated by photon-photon interactions whereby the temperature of the optical gas drops abruptly to zero. Our experiments in various configurations of coupled multicore nonlinear waveguide arrangements illustrate how light undergoing expansion-induced cooling can be channelled from arbitrary input states into the fundamental mode with near-unity efficiency. We show that the stability of the post-expansion state is ensured through an irreversible process of energy conversion. The all-optical thermodynamic phenomena explored in this study may enable innovative techniques where various uncorrelated but identical sources are merged into a unified spatially coherent state, offering a route for direct beam combining.

Employing the nonlinear self-action of light to deliberately steer and shape optical beams has been a long-standing goal in photonics[1]. Beyond the well-known examples of soliton formation[2–7], self-focusing[8–11], and filamentation[12], such scenarios often involve optical environments supporting a large number of modes, where nonlinearity enables light to dynamically explore a multitude of degrees of freedom[13–15]. Yet, as the number of modes increases, a system's nonlinear response tends to exhibit rapidly increasing complexity, often culminating in a chaotic behavior[16,17]. In this regard, notions from statistical mechanics were recently introduced in the optical domain via which such ergodic processes can be systematically understood and judiciously controlled through the lens of optical thermodynamics[18–25]. Supported by a resurgence of interest in multimoded nonlinear systems, this line of research unveiled a largely uncharted domain of optical physics where thermalization phenomena can readily manifest in entirely photonic settings[20,26–28], while simultaneously shed light on new processes in nonlinear optics such beam self-cleaning in multimode fibers[29–31]. In this context. recent experimental advancements have demonstrated thermalization of linear and orbital angular momenta to Rayleigh-Jeans equilibria in nonlinear multimode fibers[32–35], and the first observation of negative temperatures in time synthetic photonic mesh lattices[36,37]. Meanwhile, on the theoretical front, several intriguing possibilities have been proposed, such as the prospect for all-optical Carnot cycles[20] and non-Hermitian thermalization[38] as well as new methodologies to understand electrodynamic pressures at optical thermal equilibrium[39,40].

Within the framework of optical thermodynamics, one of the key quantities describing the state of a photon gas is its optical temperature[20]. Along with its corresponding chemical potential $\mu$, it dictates how the power/energy is statistically distributed among modes in a manner that maximizes the system's entropy in accord with the second law of thermodynamics[41]. Specifically, in transmission settings, the optical temperature inherently characterizes the quality of an optical beam, given the absence of statistical correlations between the modal phases at thermal equilibrium. In general, high optical temperatures indicate that the beam comprises of a great number of modes that adversely affect its coherence properties[42]. At the other extreme, a zero temperature implies that the optical power has condensed fully into the ground state whereby nonlinear mode mixing effects are substantially reduced, in analogy to a zero-temperature gas whose molecules remain almost motionless. Clearly, the prospect of optical "cooling" to significantly enhance the spatial coherence of a beam remains of paramount importance. On other occasions, there is a need to generate a coherent output by injecting light into a multimode system through a single-mode port. In the linear domain, power flowing from a single waveguide into a multi-core array results in multimode propagation and a progressive reduction of spatial coherence. At this point, a question arises as to whether a thermodynamic approach can be deployed to overcome such a fundamental linear process. If so, will it be possible to coherently couple all modes and steer power into the fundamental mode by collapsing the optical gas into a zero-temperature state? In other words, can we effectively counteract the loss of coherence, resulting from guided linear propagation, through optical cooling during a nonlinear beam expansion?

In this study, we experimentally demonstrate a Joule-Thomson (JT) expansion process by means of which the optical temperature of the photon gas can abruptly drop very close to zero (Fig. 1a, b). This effect is mediated by purely photon-photon interactions and is a direct byproduct of nonlinearity. As we will see, unlike what happens during a sudden expansion of a dense non-ideal molecular gas (Fig. 1a), here, cooling is induced by transforming the nonlinear interaction energy of the photon gas $H_{NL}$ into the kinetic component $U$ of its ground state (Fig. 1b). Yet, in both physical contexts, the end result is the same: a system initiated in an optically "hot" state (Fig. 1d) can experience a sudden drop of its temperature, transitioning into a much "colder" state (Fig. 1e) where mostly the fundamental mode is populated, through a controlled process of energy conversion (Fig. 1c). Based on optical thermodynamic principles, we establish a comprehensive theoretical framework that accurately predicts and describes the physics of this intriguing phenomenon in an all-optical context. In a series of experiments carried out with intense laser pulses in evanescently coupled waveguide arrangements[**Error! Reference source not found.**], we demonstrate up to 78% effective power transfer into the fundamental mode from a single-site excitation, with a theoretical peak efficiency of up to 90% in the absence of losses. Our observations provide a versatile approach for faithfully exciting the ground state of any arbitrary multicore geometry from a single site, while requiring minimal knowledge of the system's specific structure and properties. In a universal manner, the stability of the post-expansion state is ensured by the entropic irreversibility of the JT process. We would like to emphasize that what is observed in this study is a consequence of an irreversible nonlinear process which can be interpreted as an analogous to a thermodynamic effect and, as such, is fundamentally distinct from self-focusing or soliton formation effects. These issues are rigorously highlighted in the subsequent sections as well in the Supplementary.

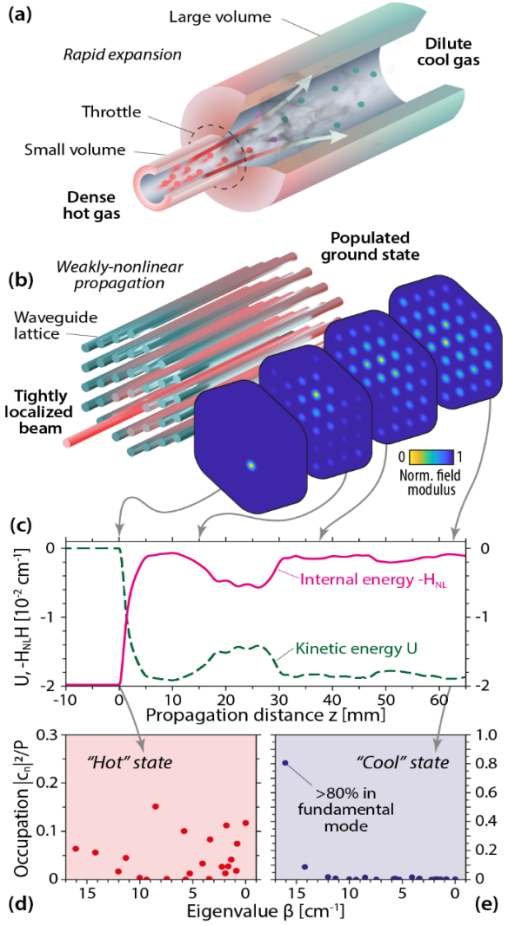

**Figure 1: Optical Joule-Thomson expansion effects.** **(a)** As a dense non-ideal molecular gas exits a narrow "throttle" and expands into a larger volume, its particles will overcome intramolecular forces at the expense of their kinetic energy, resulting in a sharp drop of temperature. **(b)** An analogous JT expansion is expected in an optical setting when an initially confined wave packet abruptly spreads across a nonlinear highly multimode system. Under the action of nonlinearity, light is funneled towards the lowest modes regardless of the specific system geometry or excitation conditions. **(c-e)** In particular, single-site excitations corresponding to optically "hot" states comprising contributions from essentially all modes may be efficiently channeled into the ground state. Note that, in line with the correspondence between wave optics and quantum mechanics, the ground state here is the state with the highest eigenvalue. To preserve intuition, the β-axes are plotted with increasing values to the left.

## Experimental platform

In addition to establishing a theoretical framework for the JT process, as detailed in the following sections, the experimental observation of the expansion effect poses a unique challenge: extracting and precisely quantifying the modal occupancies at the output of a discrete multicore array of arbitrary geometry. Here, we employ a wide range of laser-inscribed photonic lattice configurations in fused silica[43]. As illustrated in Fig. 2, specific sites of the waveguide arrays are selectively excited with intense 210 fs laser pulses with a maximum energy of 7 mJ at a wavelength of 800 nm. The light emerging from the system at the rear facet of the sample is subsequently imaged onto a pair of CMOS cameras. This enables the simultaneously recording of both the near- and far-field intensity distributions, allowing for the numerical reconstruction of the full amplitude and phase distributions of the complex optical field via the Gerchberg-Saxton algorithm[44,45] (see Methods). In turn, the modal occupancies are calculated by projection onto the sets of eigenmodes of the respective lattices, as obtained from their tight-binding Hamiltonian (see Supplementary S8.).

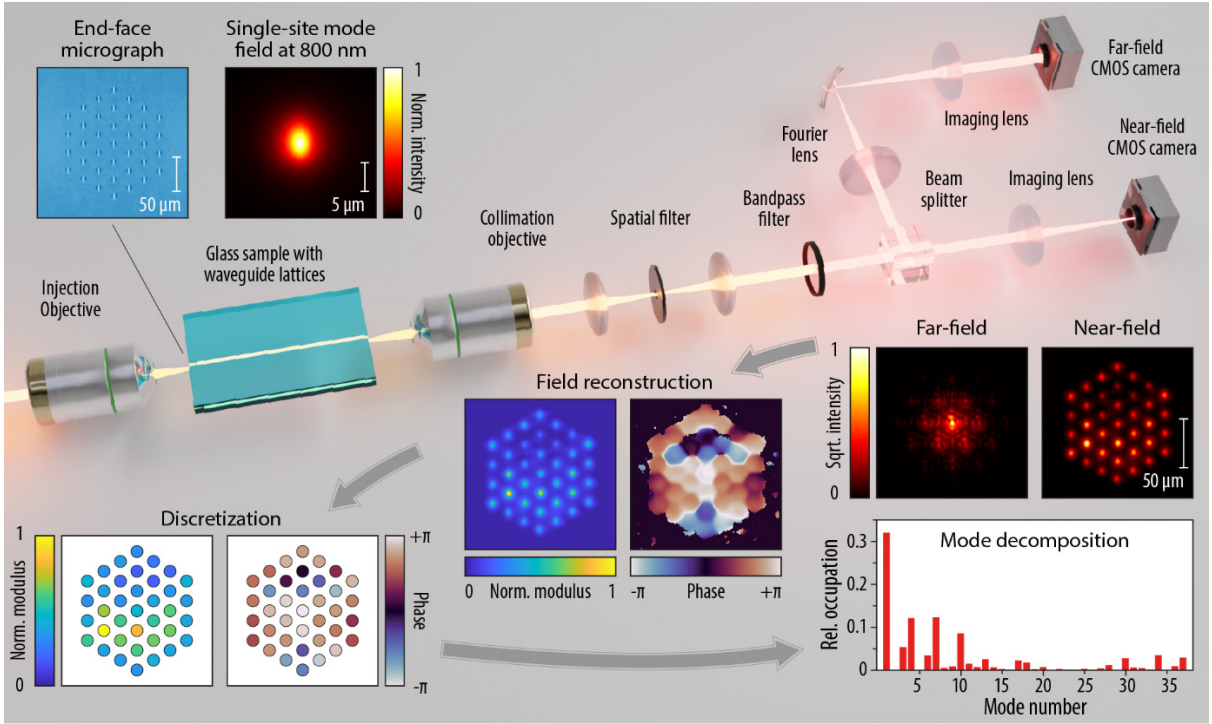

**Figure 2: Experimental setup.** Optical Joule-Thomson dynamics are investigated by injecting intense laser pulses into individual sites of a waveguide lattice embedded in a fused silica glass sample. The intensity distribution at the output facet is collimated and filtered both spatially and spectrally to obtain low-noise images in both near- and far-field in two separate beam lines. Based on the simultaneous observation of real- and Fourier-space intensity images, we numerically reconstruct the amplitude- and phase distribution of light after undergoing the JT expansion process. Finally, the modal content is calculated by projecting the discretized reconstructed wave packet onto the set of lattice supermodes.

## Theoretical description

Figure 1b illustrates a photonic Joule-Thomson expansion in a nonlinear multicore array. This process involves two distinct nonlinear optical lattices, where the smaller one (that may involve as few as one site) serves as a "throttle" to constrict the influx of light into a much larger optical system comprising $M$ cores, supporting an equal number of supermodes. In this system, the local field amplitudes $a_j$ (at site $j$) evolve along the axial distance $z$ according to

$$-id_z a_j = H a_j + |a_j|^2 a_j \tag{1}$$

where the intensity-dependent term $|a_j|^2 a_j$ results from Kerr nonlinearity and the Hamiltonian operator $H$ corresponds to the coupling matrix of the lattice configuration. In general, the field amplitudes can be expressed through the system's linear eigenstates, i.e., $a_j(z) = \sum_n^M c_n(z) e^{-i\epsilon_n z} u_{n,j}$, where $u_{n,j}$ represents the field profile of supermode $n$, while $\epsilon_n$ its corresponding eigenvalue. As a result of nonlinear four-wave mixing, the complex modal occupations $c_n(z)$ will vary along $z$ in a chaotic manner. This in turn introduces an ergodic response, enabling light to exhibit a behavior akin to that of an ideal photon gas evolving towards its thermal equilibrium.

Within a thermodynamic framework, a photon gas description of light emerges when considering the two invariants of motion associated with Eq. 1 in such conservative systems. These correspond to the optical power $P = \sum_n |a_j(z)|^2 = \sum_n |c_n(z)|^2$ conveyed in the lattice and its total Hamiltonian $H_{tot} = -U + H_{NL}$, where $U = -\sum_n \varepsilon_n |c_n(z)|^2$ and $H_{NL} = \sum_j \frac{1}{2} |a_j|^4$ are associated with the "kinetic" and "potential" energy components respectively. Provided nonlinearity is weak, the total Hamiltonian will be dominated by the kinetic component $U$, as one would expect for an ideal photon gas. In this regime, $U \simeq H_{tot}$ plays the role of the second invariant of the evolution equation. Under the constancy of $U$ and $P$, light will eventually reach equilibrium by thermalizing into a Rayleigh-Jeans (RJ) distribution[22,40], dictated by the $z$-averaged modal occupancies

$$\langle |c_n|^2 \rangle = -\frac{T}{\varepsilon_n + \mu} \tag{2}$$

an expression characterized by the optical temperature $T$ and chemical potential $\mu$. The theory of optical thermodynamics provides rigorous and unique criteria to determine the RJ equilibria from arbitrary initial conditions, thus associating the intensive quantities $(T, \mu)$ with the effective invariants of the system $(P, U)$ through a universal "equation of state". The nature of the optical temperature is further discussed in Supplementary S3.

In optical thermodynamics, the thermal state of an isolated microcanonical system is expected to remain invariant within the RJ regime. Therefore, a cooling process would only be feasible when the ideal photon-gas conditions are violated. In general, one would ideally seek to cool down a photon gas to a zero-temperature state, where the kinetic energy $U$ attains its minimum value, $U_{\min} = -\varepsilon_M P$ where $\varepsilon_M$ denotes the eigenvalue of the ground state. In this limit, light will reside permanently and exclusively within the fundamental mode, in complete analogy to a zero-temperature molecular gas with vanishing kinetic energy. Clearly, an optical system could experience such a cooling process if a controlled exchange between $U$ and $H_{NL}$ is allowed to take place. In this respect, a new nonlinear regime must be identified where light could exhibit a behavior akin to a non-ideal photon gas.

In classical thermodynamics, a sudden expansion can precipitate an irreversible exchange between the kinetic and intramolecular energy components of a non-ideal Van der Waals gas, overcoming the intramolecular forces at the expense of its kinetic energy. In optics, an analogous effect can be observed when a photon gas transitions abruptly from an initially confined multimoded system into a significantly larger one at moderately high-power levels (Fig. 1c). During this process, conservation of the Hamiltonian energy $H_{tot}$ mandates that the combined potential and kinetic components before expansion must be equal to the total energy after expansion, i.e., $-U_{\mathrm{L}} +$

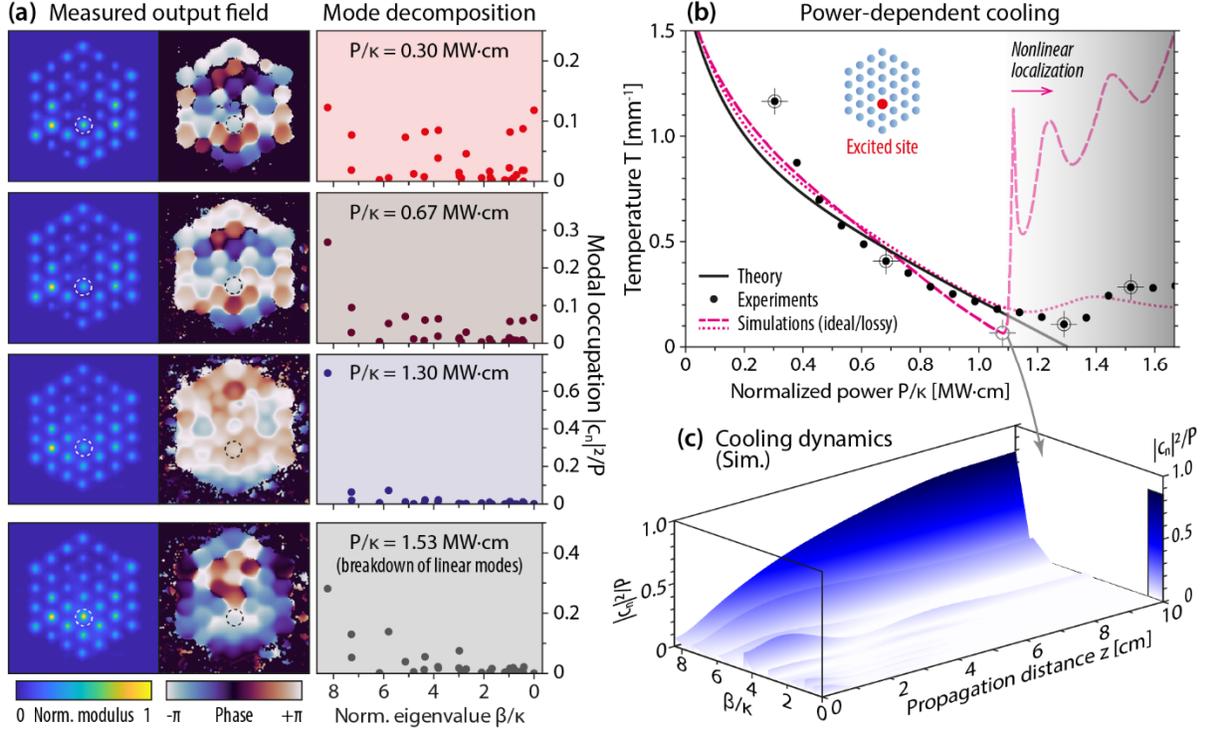

**Figure 3: Joule-Thomson expansion in a triangular lattice.** (a) Amplitude- and phase profiles (left) and corresponding modal content (right) observed at the output facet of the multicore lattice for different normalized excitation powers P/κ. Despite a fixed placement of the initial excitation (indicated by dashed circles), the initially broad population of most lattice modes (P/κ = 0.30 MW · cm) gradually cools down until approximately 70% of light is channeled into the fundamental mode at P/κ = 1.30 MW · cm. Eventually, the onset of nonlinear localization at even higher powers leads to a breakdown of the linear mode decomposition (see bottom row for P/κ = 1.53 MW · cm). (b) The optical temperatures corresponding to the entire set of measurements (black dots) are compared to BPM simulations in a lossless lattice (magenta solid line) and the lossy one used in our experiments (magenta dashed line, 0.3dB/cm), respectively. The theoretical curve (black solid line) follows Eq. 3 in physical units (see Supplementary S2.). Note that even in the presence of losses, the JT process still manifests itself through thermalization to a nearly zero temperature. The data points corresponding to the panels in (a) are indicated by black crosshairs. (c) Simulated JT cooling dynamics (BPM) in a lossless lattice are shown for the minimum of the magenta solid curve in (b), with approximately 80% of power irreversibly funneled in the fundamental mode after a propagation distance of 10 cm

$H_{\text{NL,L}} = -U_{\text{S}} + H_{\text{NL,S}}$, where the subscripts S, L denote the small (input) and large lattice respectively. In the following, we will consider the special case where the expansion is initiated from a single waveguide (the general multi-site excitation case is described in Supplementary S4). In this situation, light occupies the smallest possible spatial domain (a single site in the S lattice) and photon-photon interactions dominate. As a result, the potential energy before expansion will be $H_{\text{NL,S}} = H_{\text{NL,max}} = \frac{1}{2}P^2$, thus assuming its global maximum for any input power $P$. In this case, the pre-expansion kinetic component is $U_s = 0$.

In the linear domain, as light enters the larger system L from the single-element input of S, it will disperse across its entire cross-section. These selective excitation conditions result in multimode propagation involving almost all modes of the guide network, with a specific relative phase structure. Evidently, this scenario totally precludes the possibility for funneling all the power in the ground state, as the amplitudes of the modes will remain invariant during propagation. The situation is drastically different in the presence of nonlinearity. In this case, the potential energy will rapidly drop close to its minimum value, given by $H_{\mathrm{NL,L}} \approx H_{\mathrm{NL,min}} = \sum_j \frac{1}{2} |a_j|^4 = \sum_j \frac{1}{2} |P/M|^2 = \frac{1}{2} P^2/M$, indicating an almost uniform distribution of power amongst all lattice sites. For sufficiently large lattices ($M \gg 1$), this post-expansion potential energy $H_{\mathrm{NL,min}}$ becomes negligible. Ultimately $H_{\mathrm{NL,S}}$ will be transferred entirely to the kinetic component, i.e. $-U_{\mathrm{L}} \approx H_{\mathrm{NL,S}}$, resulting in

$$U_L = -\frac{1}{2} P^2 \tag{3}$$

Equation 3 indicates that the post-expansion kinetic energy of a non-ideal photon gas will depend exclusively on the injected power. In particular, an increase of $P$ at the input corresponds to a commensurate decrease in the post-expansion kinetic energy. This simple relation allows one to directly control the resulting optical temperature of a photon gas, and, accordingly, the degree to which light is concentrated in the lower order modes once injected from a single-site input. Most importantly, a zero-temperature state can be faithfully manifested at a certain power level that corresponds to the minimum kinetic energy of the large system, $U_{\mathrm{L,min}} = -\varepsilon_M P$, where all the optical power is expected to be funneled into the fundamental mode. By substituting this expression into Eq. 3 we find $-1/2P^2 = -\varepsilon_M P$ and consequently,

$$P = 2\varepsilon_M \tag{4}$$

Equation 4 provides a universal and lattice-independent power value at which the fundamental mode (with eigenvalue $\varepsilon_M$) is reliably excited from a single-site input via this all-optical JT expansion process.

**Experimental results**

Figure 3a depicts the first representative example of our experimental observations in a finite hexagonal domain involving $M = 37$ sites (or supermodes), built on a triangular unit cell with a nearest-neighbor coupling coefficient of $\kappa = 0.3 \ \mathrm{cm}^{-1}$. Light is injected into the lattice via a single waveguide port, a point-like excitation that simultaneously populates virtually all supermodes. A closer look at the output distributions provides valuable insights into the mechanisms at play. At low power levels, the quasi-linear diffraction pattern of the single-site input (obtained at a normalized power of $P/\kappa = 0.30 \ \mathrm{MW \cdot cm}$) displays a complex phase pattern with numerous sign flips between adjacent sites (top row of Fig.3a). In this regime, the expansion process has a negligible impact on the photon gas, as illustrated in the red shaded panel of Fig.3a where the lattice supermodes retain a random-like distribution. As power increases (second row of Fig.3a), the first indications of optical cooling emerge, with optical power progressively moving towards lower order modes.

Further increase in the injected power (third row of Fig.3a) results in a more homogenized light distribution across the waveguide array in both phase and amplitude. At this particular power level ($P/\kappa = 1.3 \ \mathrm{MW \cdot cm}$), the output distribution approximately conforms with the smooth amplitude envelope and the flat phase of the ground state, indicating that the power now almost exclusively resides in the fundamental mode. From experimental data, and our theoretical analysis, we estimate that, post-expansion, the photon gas in the larger array will eventually relax at thermal equilibrium to a Rayleigh-Jeans distribution with an optical temperature $T = 0.11 \mathrm{m}^{-1}$ and a chemical potential $\mu = -156\mathrm{m}^{-1}$. For the triangular lattice used in our experiments, the power corresponding to the JT expansion process (in actual units) can be obtained directly from the renormalized Eq. 4, and is given by $P = 10\kappa c\epsilon_0 n_0 A_{eff}/k_0 n_2$, where $A_{eff}$ is the effective area of the waveguide mode and $n_2$ the silica nonlinear coefficient. A derivation of this expression based on physical units can be found in Supplementary S1. For the parameters used in our experiment, we predict a critical power value of 390kW which approximately matches the experimental observations.

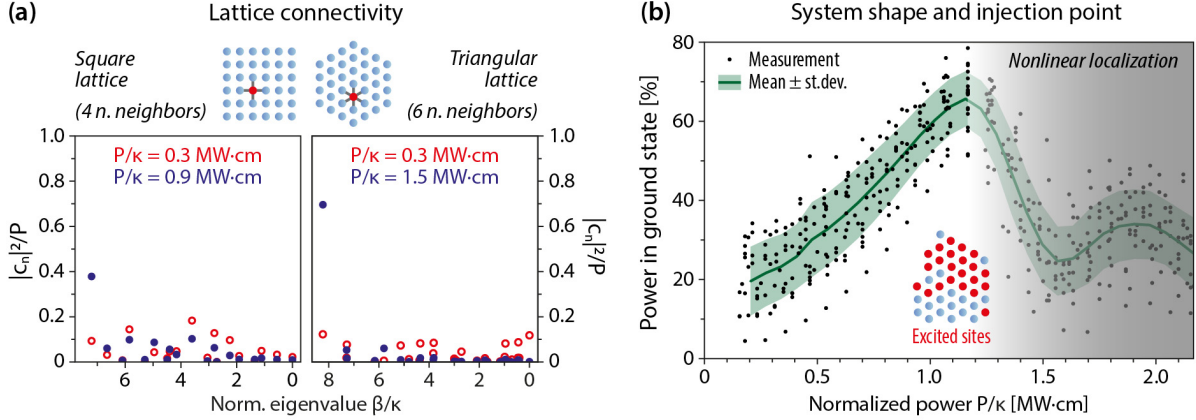

**Figure 4: Impact of lattice connectivity, system shape and excitation position.** (a) The JT expansion mechanism is universal and fundamentally independent of the system characteristics. Nonetheless, the higher threshold for nonlinear self-localization in more tightly connected networks allows for lower temperatures to be reached, as shown in our experiments comparing a square lattice (left) to the triangular lattice (right). Featuring only four nearest-neighbor sites, the minimum post-expansion temperature occurs at slightly lower powers in the square versus the triangular lattice, resulting in correspondingly lower ground state populations (37% vs 69 %). (b) The specific placement of the single-site initial excitation has only a minor influence on the expansion process, as shown here for 22 different excitations in an irregularly shaped triangular lattice, which lacks any geometrical symmetries. The ensemble average and standard deviation are indicated by a solid line and its surrounding shaded region, whereas individual measurements are shown as black dots. Details on the thermalization dynamics in the three lattices are shown in Extended Data Figure XD3

The reconstructed modal distributions at the output allow us to track the post-expansion kinetic energy $U_L$ and the optical temperature $T$ of the photon gas over a wide range of input powers. In Fig. 3b, the observed dependence of $T$ on the excitation power (data points) is compared with Beam-Propagation-Method (BPM) simulations (magenta lines) and theory (black line). Overall, the response detailed in Fig. 3b closely aligns with the monotonic dependence of the post-expansion kinetic energy $U_L$ on the optical power $P$, as described by Eq. 3, which results in a gradual decrease of the optical temperature. In our experiments we found that even in the presence of propagation losses (0.4 dB/cm in our sample), a near-zero temperature expansion results in a fundamental mode occupancy of approximately 70% in the transmitted power, while our simulations suggest that in a lossless environment, a conversion efficiency of nearly 90% would be readily attainable from a single-site excitation without any need for beam shaping (see Extended Data Figure XD1a.). Figure 3b represents the JT response of this system when losses are also taken into account (dashed magenta curve). Evidently, while the thermodynamic theory of light relies on the conservation of optical power, here losses have a minimal impact on the expansion process as cooling occurs rapidly, resulting in only 10%-20% loss of efficiency in generating a maximally coherent fundamental mode output.

Beyond the JT limit, as the power increases, we experimentally observe that the photon gas model suddenly breaks down and the projection of the reconstructed field distribution onto the linear modes of the system becomes increasingly unreliable (last row of Fig.3a). In the BPM simulations, this coincides with an apparent sharp increase of the optical temperature (as seen in the solid red curve of Fig.3b) that corresponds to a soliton self-focusing collapse. This marks the final departure from the JT regime.

In general, while Eq. 3 formally relies on an irreversible spreading of light across the entirety of the transverse cross-section of the lattice, this assumption is not particularly restrictive. Absent certain pathological cases such as compact states[Error! Reference source not found.] or disorder-induced Anderson localization[46], such dispersive spreading is generally observed in any linear network of coupled identical photonic elements. In particular, once light undergoes irreversible expansion it will remain indefinitely in the fundamental mode. Our experiments can only provide information at the output of the multicore array; therefore, we validate this aspect via tight-binding simulations which illustrate the full z-evolution of the kinetic energy $U$ (see Extended Data Figure XD2a) at such elevated power levels. Evidently, the nonlinear potential energy $H_{NL}$, calculated along the axial direction z, drops and remains very close to zero while being transferred fully and irreversibly to the kinetic component. These results confirm the theoretical assumptions of Eq. 3, where near- perfect energy conversion was assumed.

To illustrate the universality of this process, we extend our investigation beyond the triangular lattice considered so far. In doing so, we employ both square and irregularly shaped multicore geometries (see Fig. 4 and Extended Data Figure XD2). Our simulations indicate that while the minutiae of the expansion dynamics naturally depend on the structure of the lattice, the JT expansion close to the zero-temperature state can still occur in all types of lattices, thereby enabling ground-state conversion with efficiencies of 79% in the square lattice and 91% in the irregular triangular lattice for lossless conditions, respectively. Note that these lattices feature four and six nearest neighbors per site, respectively, and the correspondingly higher thresholds of soliton formation further boost the JT efficiency in more tightly connected systems, as qualitatively confirmed by our measurements (Fig. 4a). The absence of geometrical symmetries in the irregular lattice further favors the expansion process versus self-localization, allowing for such high efficiencies.

Finally, we experimentally investigate how the input site location affects the mode conversion efficiency during the JT process. By virtue of the general nature of the JT expansion, the interaction-driven population transfer to the ground state occurs irrespective of the specific placement of the initial confined wave packet, or the geometry of the lattice itself. In theory, the post-expansion temperature of the photon gas should be entirely agnostic to the input location. As the ensemble average over measurements at 22 different distinct injection points in an irregularly shaped triangular lattice domain demonstrates, this robustness impressively carries over to finite propagation lengths in lossy arrangements, as the efficiency follows the same qualitative trajectory regardless of the injection site.

**Discussion**

In conclusion, we have demonstrated, for the first time, photon-photon thermodynamic Joule-Thomson expansion processes by means of which light can be effectively funneled into the ground state of a nonlinear highly multi-mode system in a self-organized fashion. Our experiments and simulations show that over 75% ground state conversion efficiencies can be readily achieved, irrespective of the geometry of the system or the location of the input. As indicated in our study, the mechanism of the observed JT expansion relies on the energy exchange between the nonlinear component of the Hamiltonian and the "kinetic energy" associated with the fundamental mode. In this respect, it is exclusively a thermodynamic effect and carries no connection to the self-focusing action of nonlinearity. To further substantiate this claim, we consider an ideal JT expansion in a fully linear lattice where nonlinearity is present only at the injection site of the smaller lattice (see Supplementary S5). Even in this scenario where the larger lattice is entirely linear, the JT effect still manifests itself with high conversion efficiency.

Another important aspect is the type of nonlinearity itself, which is responsible for the observed light JT expansion. The fused silica platform employed here exhibits a focusing Kerr nonlinearity. One should expect a converse effect in defocusing media, where light expansion will instead lead to "optical heating" ($T \to 0^-$). This possibility is illustrated in Supplementary Figure S4 where the power injected into a square lattice is now predominantly funneled into the highest-order mode (characterized by a staggered eigenmode profile) when again the input power is close to the value predicted by Eq. 4 (where now $\varepsilon_M$ is replaced by $\varepsilon_1$, being the eigenvalue of the highest order mode), while the kinetic energy rapidly increases to $U_L = 1/2P^2$ (with a positive sign).

The thermodynamic effects investigated here can lead to novel methodologies where multiple inputs are employed simultaneously to drive a multimode arrangement into a single spatially coherent state, thus providing a pathway for optical beam combining. Simulations supporting these arguments are presented in Supplementary S4 where a dual-input scenario is investigated. Our results could also impact other areas of science and technology involving nonlinear classical bosonic arrangements, such as Bose-Einstein condensates[47,48], superconducting platforms[49] and magnonic systems[50].

.

**Methods**

**Waveguide fabrication and nonlinear characterization**

Our photonic lattices were fabricated with the femtosecond laser direct writing technique. Specifically, ultrashort pulses from a frequency-doubled fiber amplifier system (Coherent MONACO, 270 fs pulse length, repetition rate 333 kHz, carrier wavelength 517 nm) were focused into the bulk of a 100 mm long fused silica sample. A precision positioning system (Aerotech ALS150) was used to translate the sample with respect to the focal spot, resulting in single-mode waveguides with propagation losses of approximately 0.4 dB/cm and mode field diameters of approximately 13µm × 15µm at the probe wavelength of 800 nm.

For our experiments, individual lattice sites were excited by ultrashort pulses from a titanium sapphire chirped pulse amplifier (Coherent ASTRELLA, 210 fs pulses, repetition rate 1 kHz, carrier wavelength 800 nm, maximum pulse energy 7 mJ with approximately 1% pulse-to-pulse fluctuations). After propagation through the sample, the output intensity distribution was collimated by a $4\times$ microscope objective ($NA = 0.1$) and subjected to spatial as well as spectral filtering to reduce the influence of scattered-light background and constrain the bandwidth to approximately 10 nm to aid phase reconstruction. To this end, the beam was subsequently passed through a 50:50 beam splitter to allow for the simultaneous capture of the near field and Fourier domain images on a pair of CCD cameras (Basler acA1920-155um). The images were averaged over approximately 100 pulses. The pulse spectra measured before and after propagation in the sample are shown in Supplementary Figure S10.

**Reconstruction of the electric field and mode occupations**

To extract the contributions of the individual lattice modes to the total light distribution at the output facet of the sample, the full field needs to be determined. While the field modulus is obtained readily enough from the intensity amplitudes, additional information is required to access the phase. To this end, we employ the Gerchberg-Saxton-Algorithm[Error! Reference source not found.], in which the square root of the measured near field intensity, together with a random phase distribution, serves as first guess that is refined by going back and forth between the real space- and Fourier domain, in which the square root of the observed far field intensity is injected in each iteration cycle. The algorithm typically converges after a few hundred iterations, as tracked by the far field error

$$\sigma = \sum_{x',y'} \left| |E_{\text{Fourier}}(x',y')|^2 - \left|\tilde{E}_{\text{Fourier}}(x',y')\right|^2 \right|^2$$

that compares the measured intensity $|E_{\text{Fourier}}|^2 = I_{\text{Fourier}}$ to the retrieved Fourier domain intensity $\left|\tilde{E}_{\text{Fourier}}\right|^2$. To exclude influences of the initial random phase, the retrieval routine was repeated 10 times for every measurement pair, each time with a different random initial phase, to verify reproducible convergence of $\sigma$ to the same value.

To retrieve the modal occupancies, the retrieved complex output field was discretized into a state vector $a$ by averaging amplitude and phase around the positions of each waveguide with index $j$ with

$$|a_j|e^{i\varphi_j} = |E_{\text{near}}(x_j, y_j)|e^{i\varphi(x_j, y_j)} .$$

In turn, the full set of orthogonal eigenmodes $u_n$ and corresponding eigenvalues $\epsilon_n$ of the discrete lattices were calculated from the respective tight-binding Hamiltonians to obtain the modal occupations $c_n$ via the projection $c_n = \langle a|u_n \rangle$.

**Extended Data**

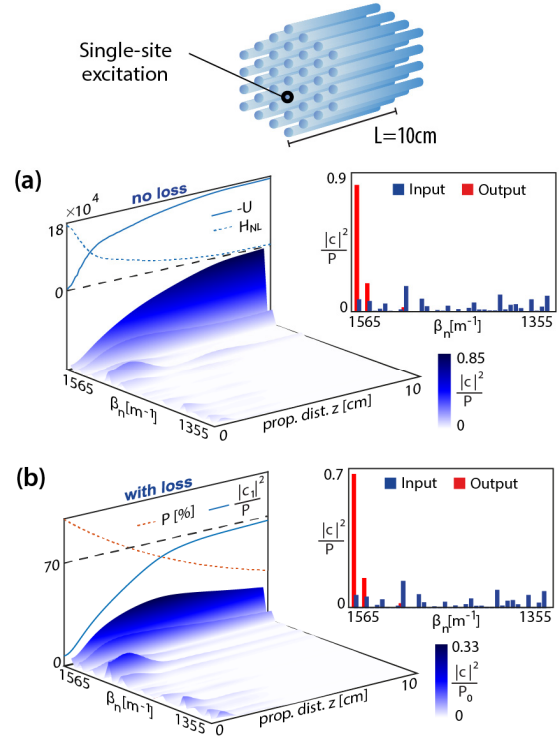

**Extended Data Figure XD1: Thermalization in the presence of losses: BPM simulations (a)** In an ideal loss-less system, the overall power is conserved, and the total power contained in the fundamental mode steadily grows during the cooling process. (b) The presence of losses introduces a global power decay (dashed red line), but the relative fraction of power contained in the fundamental mode (solid blue line) at a given propagation distance nevertheless increases as thermalization proceeds. As shown in Fig. 3b, even in the presence of significant propagation losses attenuation 0.3 dB/cm, a value beyond the typical losses of 0.4dB/cm for laser-written waveguides in fused silica, the efficiency of the JT power transfer to the ground state only decreases by approximately 10%.

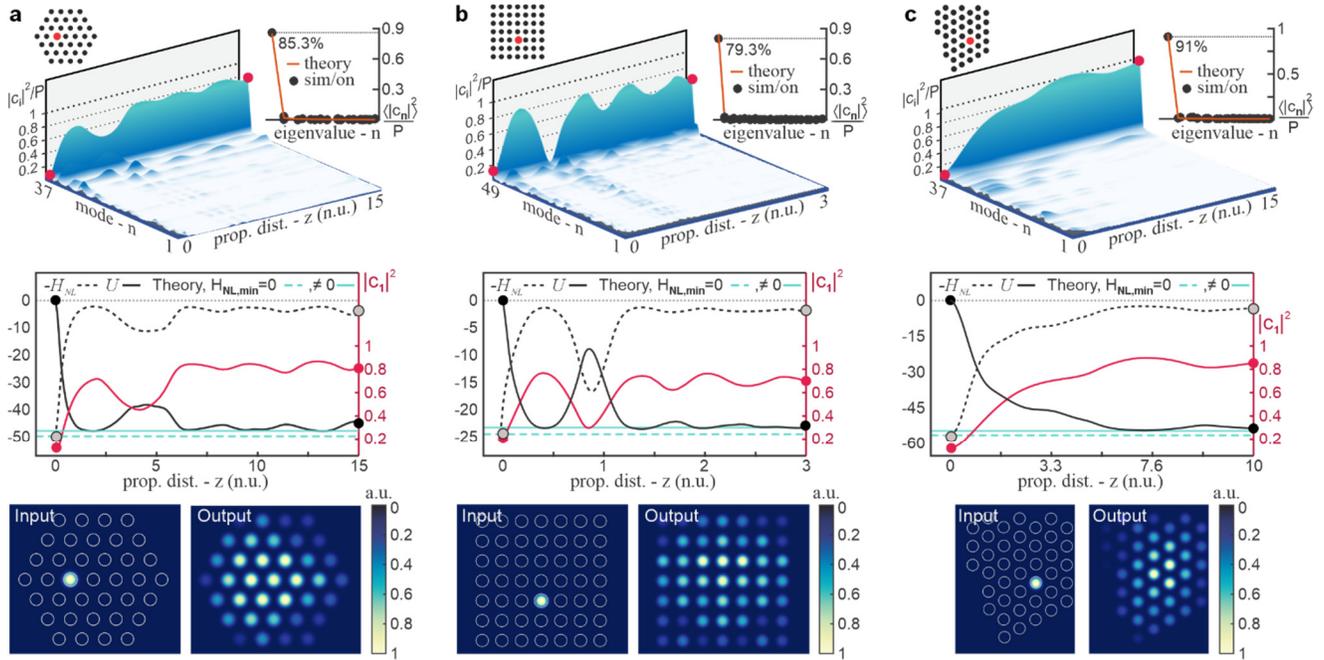

**Extended Data Figure XD2: Nonlinear dynamics in the (b) square and (a,c) triangular lattices: tight-binding simulations.** In all lattices, light was injected from a single site with peak power favoring a JT irreversible expansion, just below the threshold of self-focusing collapse. The average (Rayleigh-Jeans) conversion efficiency to the fundamental mode is 85.3%, 79.3% and 91% for the, triangular (hexagonal), square, and triangular (irregular) lattice, respectively. An irreversible exchange of energy is observed between $U$ and $H_{NL}$, matching almost perfectly with theoretical prediction. The teal dashed line represents Eq. 4 while the teal solid line includes a correction term, $U_L = -\frac{1}{2}P^2 + H_{NL,min}$.